\documentclass{icrc2009}

\usepackage{graphicx}   
\usepackage[caption=false]{caption}    
\usepackage[font=footnotesize]{subfig} 
\usepackage{fixltx2e}
\usepackage{url}

\newcommand{\shorttitle}[1]%
{\markboth{Proceedings of the 31\MakeLowercase{$^{st}$} ICRC, {\L}\'{o}d\'{z} 2009}{#1} }
\newcommand{\etal}{\MakeLowercase{\textit{et al. }}} 

\begin{document}
\title{A TeV source in the 3C 66A/B region}

\author{\IEEEauthorblockN{Manel Errando\IEEEauthorrefmark{1},
			  Elina Lindfors\IEEEauthorrefmark{2},
                          Daniel Mazin\IEEEauthorrefmark{1},
                           Elisa Prandini\IEEEauthorrefmark{3} and
                           Fabrizio Tavecchio\IEEEauthorrefmark{4}}
                           for the MAGIC Collaboration
                            \\
\IEEEauthorblockA{\IEEEauthorrefmark{1}IFAE, Edifici Cn., Campus UAB, E-08193 Bellaterra, Spain}
\IEEEauthorblockA{\IEEEauthorrefmark{2}Tuorla Observatory, Turku University, FI-21500 Piikki\"o, Finland}
\IEEEauthorblockA{\IEEEauthorrefmark{3}Universit\`a di Padova and INFN, I-35131 Padova, Italy}
\IEEEauthorblockA{\IEEEauthorrefmark{4}INAF National Institute for Astrophysics, I-00136 Rome, Italy}}

\shorttitle{Errando \etal A TeV source in the 3C 66A/B region}
\maketitle

\begin{abstract}
The MAGIC telescope observed the region around the distant blazar 3C~66A for 54.2\,hr in 2007 August--December. 
The observations resulted in the discovery of a $\gamma$-ray source 
centered at celestial coordinates R.A. = $\mathbf{2}^{\mathrm{\mathbf{h}}} \mathbf{23^{\mathrm{\mathbf{m}}}} \mathbf{12}^{\mathrm{\mathbf{s}}}$ and decl.$\mathbf{=43}^{\mathbf{\circ}} \mathbf{0.'7}$ (MAGIC~J0223+430), coinciding with the nearby radio galaxy 3C~66B. 
A possible association of the excess with the blazar 3C~66A is discussed.
The energy spectrum of MAGIC~J0223+430 follows a power law with a normalization of $\left(\mathbf{1.7\pm 0.3}_{\mathrm{\mathbf{stat}}} \mathbf{\pm 0.6}_{\mathrm{\mathbf{syst}}} \right)\mathbf{\times10}^{\mathbf{-11}}$ 
TeV$^{-1}$ cm$^{-2}$ s$^{-1}$ at 300\,GeV and a photon index $\mathbf{\Gamma = -3.10 
\pm 0.31}_{\mathrm{\mathbf{stat}}}\mathbf{ \pm 0.2}_{\mathrm{\mathbf{sys}t}}$. 
  \end{abstract}


 \section{Introduction}
\label{intro}
As of today, there are 26 known extragalactic very high energy (VHE, defined
here as $E>100$\,GeV) $\gamma$-ray sources. All of them are active galactic
nuclei (AGNs) with relativistic jets. With the exception of the radio galaxy
M~87 and Cen\,A all detected sources are blazars, whose jets (characterized by
a bulk Lorentz factor $\Gamma \sim 20$) point, within a small angle ($\theta
\sim 1/\Gamma$), to the observer. The spectral energy distribution (logarithm of the observed energy density versus logarithm of the photon energy)
of AGNs typically show a two-bump structure. 
Various models have been proposed for the origin of the high-frequency bump,
the most popular being inverse Compton scattering of low-frequency photons. There
have been several suggestions for the origin of the low-frequency seed photons
that are up-scattered to $\gamma$-ray energies: they may be produced within the
jet by synchrotron radiation (synchrotron self-Compton or SSC
mechanism\,\cite{Maraschi}) or come from outside the jet (external Compton or EC
mechanism\,\cite{Dermer}). Relativistic effects boost the observed emission as the
Doppler factor depends on the angle to the line of sight. In case the jet angle
to the line of sight is large, models that depend less critically on beaming
effects are needed \cite{Tavecchio}. The VHE $\gamma$-ray emission of
AGNs might also be of hadronic origin through the emission from secondary
electrons \cite{Mannheim}. 

3C~66A and 3C~66B are two AGNs separated by just $6'$ in the sky.  3C~66B is a
large Fanaroff--Riley-I-type (FRI)
radio galaxy, similar to M~87, with a redshift of 0.0215 \cite{Stull}, 
whereas 3C~66A is a distant blazar. As pointed aout in \cite{Bramel}, the often referred redshift of
0.444 for 3C~66A is uncertain.  It is based on a single measurement of one emission
line only \cite{Miller}, 
while in later observations no lines in the spectra of 3C~66A were reported
\cite{Finke}. Based on the marginally resolved host galaxy, a
photometric redshift of $\sim 0.321$ was inferred \cite{Wurtz}.
In this paper we report the discovery of VHE $\gamma$-ray emission located $6.'1$ away from the blazar 
3C~66A and coinciding with the radio galaxy 3C~66B in 2007. Detailed results and discussion
can be found in \cite{magic3c66}.


\section{Observations and Data Analysis}
\label{analysis}

3C~66A underwent an optical outburst in 2007 August, as monitored by the Tuorla 
blazar monitoring program. The outburst triggered VHE $\gamma$-ray 
observations of the source with the MAGIC telescope following the Target of Opportunity program,
which resulted in discoveries of new VHE $\gamma$-ray sources in the past.

MAGIC has a standard trigger threshold of 60\,GeV, an angular
resolution of $ \sim 0.^\circ 1$ and an energy resolution
above 150\,GeV of $\sim 25\%$ (see \cite{crab} for details).
The MAGIC data analysis is described in detail in \cite{magic3c66,crab}.

 \begin{figure*}[htpb]
   \centerline{\subfloat{\includegraphics*[height=6.2cm]{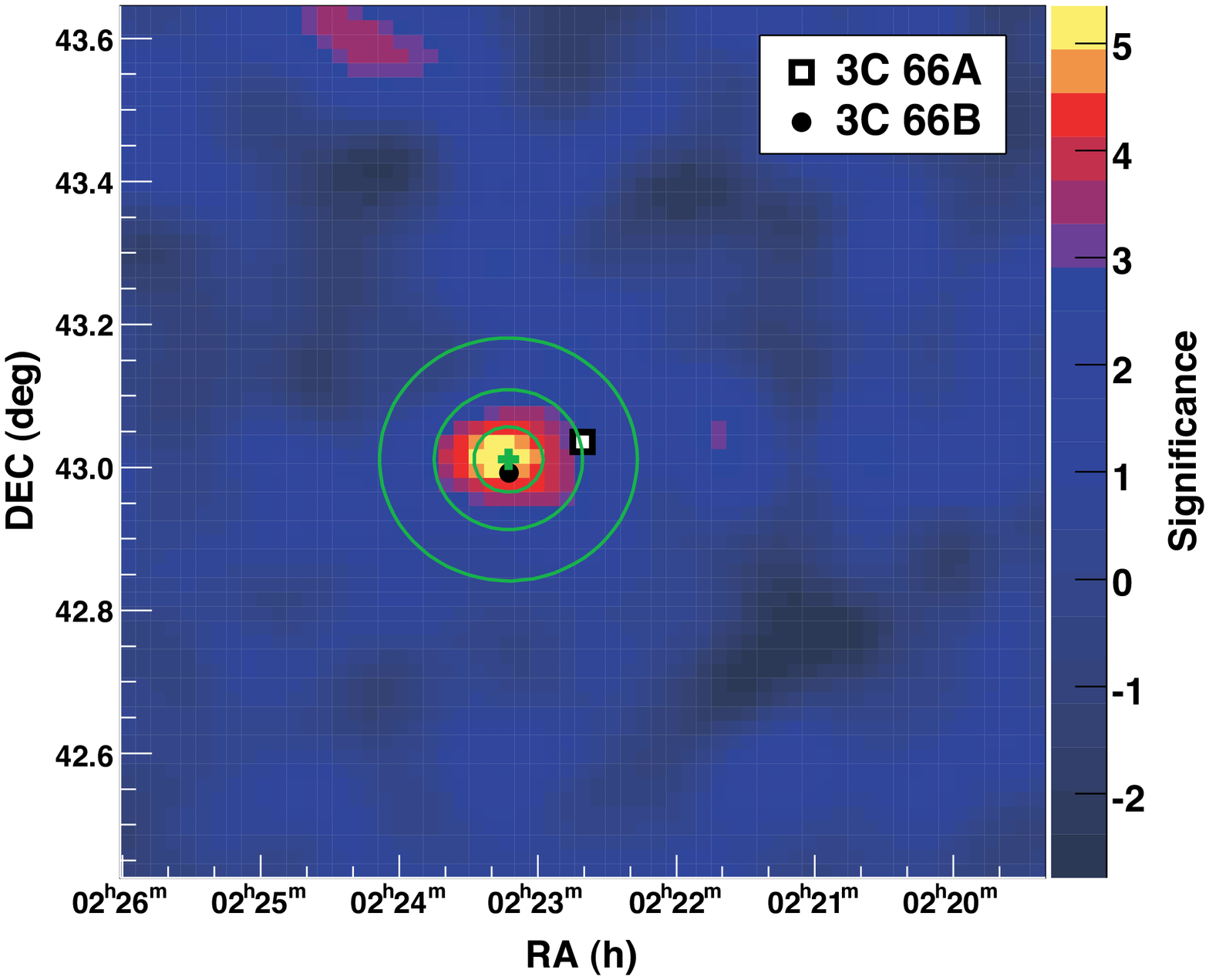}}
              \hfil
              \subfloat{\includegraphics*[height=6.6cm]{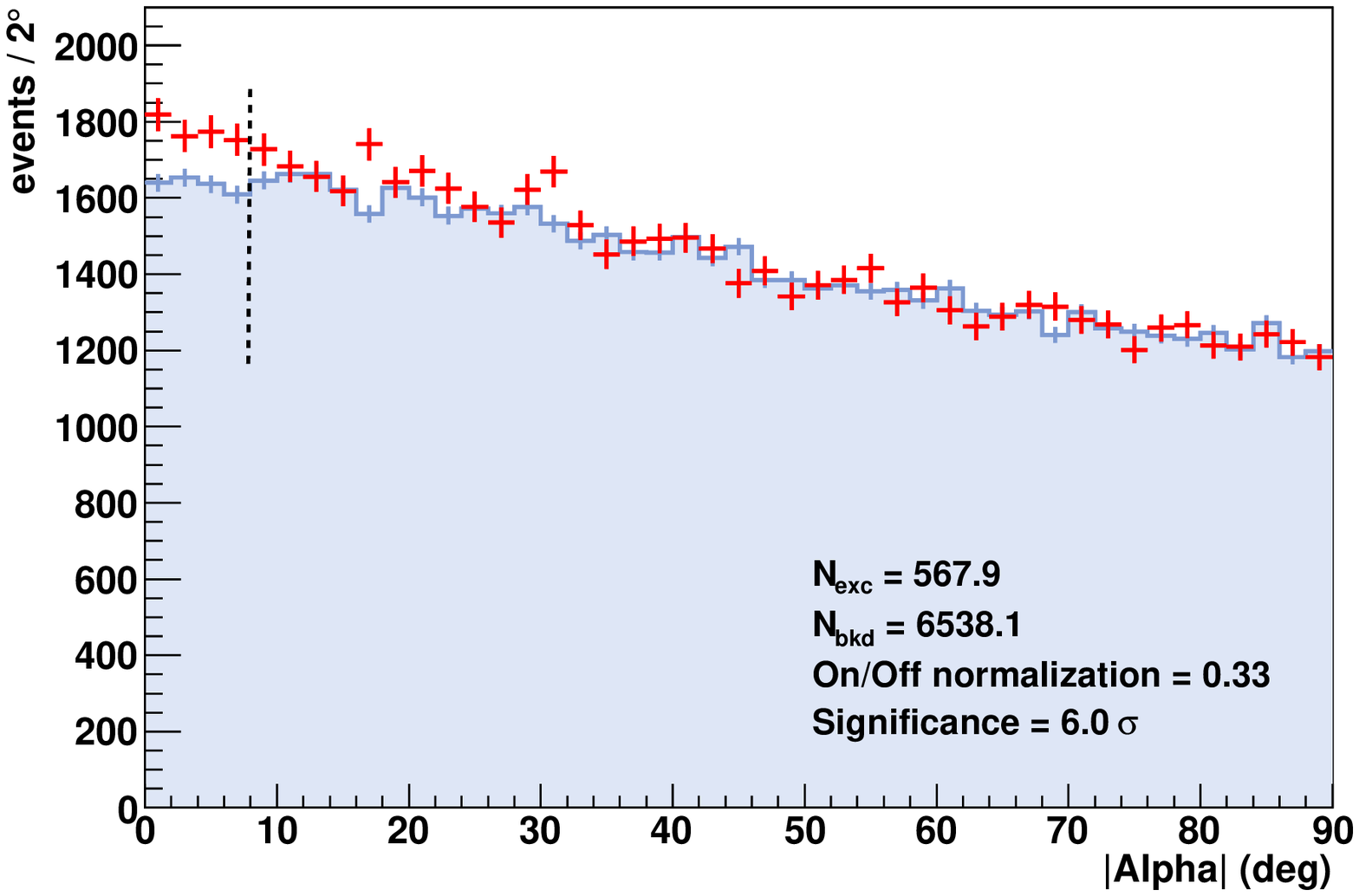}}
             }
   \caption{\textit{Left plot:} Significance map for $\gamma$-like events above 150\,GeV 
in the observed sky region. The green cross 
corresponds to the fitted maximum excess position of MAGIC~J0223+403. 
The probability of the true source to be inside 
the green circles is 68.2\%, 95.4\% and 99.7\% for the inner,
middle and outer contour, respectively. 
The catalog positions 
of 3C~66A and 3C~66B are indicated by a white square and a black dot, respectively. 
\textit{Right plot:} $\mid$Alpha$\mid$ distribution after all cuts evaluated with respect to the position of MAGIC~J0223+430. A $\gamma$-ray excess with a significance of $6.0\,\sigma$ is found, which corresponds to a post-trial significance of $5.4\,\sigma$.}
 \label{fig:skymap}
  \label{fig:alpha}
 \end{figure*}

Data were taken in the false-source tracking (wobble) mode
pointing alternatively to two different sky
directions, each at $24'$ distance from the 3C~66A catalog
position. The zenith distance distribution of the data extends from 13$^{\circ}$ to 35$^{\circ}$. Observations were made in 2007 August, September, and December and lasted 54.2\,hr, out of which 45.3\,hr passed the
quality cuts based on the event rate after image cleaning. 
An additional cut removed the events with total charge 
less than 150 photoelectrons (corresponding to an energy of $\sim 90$\,GeV)
in order to assure a better background rejection.

Just before the start of the observation campaign $\sim 5 \%$ of the mirrors on the telescope were replaced, worsening the optical point-spread function (PSF). As a consequence, a new calibration of the mirror alignment system became necessary, which took place within the observation campaign and improved the PSF again. 
The sigma of the Gaussian PSF (40\% light containment) was measured to be $3.'0$ in 2007 August 12-14, $2.'6$ in 2007 August 15-26 and $2.'1$ in 2007 September and December. 
To take this into account, data were analyzed separately for each period and the results were combined at the end of the analysis chain. However, the realignment resulted in a mispointing, which was taken care of by a new pointing model applied offline using starguider information \cite{bretz}. Considering the additional uncertainty caused by the offline corrections, we estimate the systematic uncertainty of the pointing accuracy to be $2'$ \cite{drive}. 


\section{Results}
\label{results}

Figure \ref{fig:skymap} (left plot) shows a significance map produced from the signal and background maps, 
both smoothed with a Gaussian of $\sigma=6'$ 
(corresponding to the $\gamma$-PSF), for photon energies between 150\,GeV and 1\,TeV. 
Loose background rejection cuts are applied  to keep a large number of gamma-like events.
The center of gravity of the $\gamma$-ray emission is derived from Figure~\ref{fig:skymap}
 by fitting a bell-shaped function of the form
\begin{equation}
F(x,y) = A \cdot \exp \left[ -\frac{(x-\bar{x})^2+(y-\bar{y})^2}{2\sigma^2}\right]
\label{gaus}
\end{equation}
for which the distribution of the excess events is assumed to be rotationally symmetric, i.e., $\sigma_x=\sigma_y=\sigma$. 
The fit yields reconstructed coordinates of the excess center of R.A. = $2^{\mathrm{h}} 23^{\mathrm{m}} 12^{\mathrm{s}}$ and decl.$=43^{\circ} 0.'7$.
The detected excess, which we name MAGIC~J0223+430, is $6.'1$ away from the catalog position of 3C~66A, 
while the distance to 3C~66B is $1.'1$. We made a study to estimate 
statistical uncertainty of the reconstructed position. 
The probabilities are shown in Figure~\ref{fig:skymap} by the green contours 
corresponding to 68.2\%, 95.4\%, and 99.7\% for the inner,
middle, and outer contour, respectively.
Using this study we find that the measured excess coincides with the catalog position of 3C~66B.
The origin of the emission from 3C~66A can be excluded with a probability of 95.6\% ($2.0\,\sigma$). Linearly adding the systematic uncertainty of the pointing for this data set ($2'$, see above), i.e., shifting the excess position by $2'$ toward the catalog position of 3C~66A, the exclusion probability is 85.4\% ($1.5\,\sigma$).

To calculate the significance of the detection, 
an \textsc{$\mid$Alpha$\mid$} distribution was
produced (see Figure~\ref{fig:skymap}, left), where \textsc{Alpha} is the angle between the major axis of the
shower image ellipse and the source position in the camera. 
A signal of $6.0\,\sigma$ significance (pre-trial, at the position of 3C~66B) 
and $5.4\,\sigma$ (post-trial, using 30 independent trials) has been calculated.

 \begin{figure*}[htpb]
   \centerline{\subfloat{\includegraphics*[height=10cm]{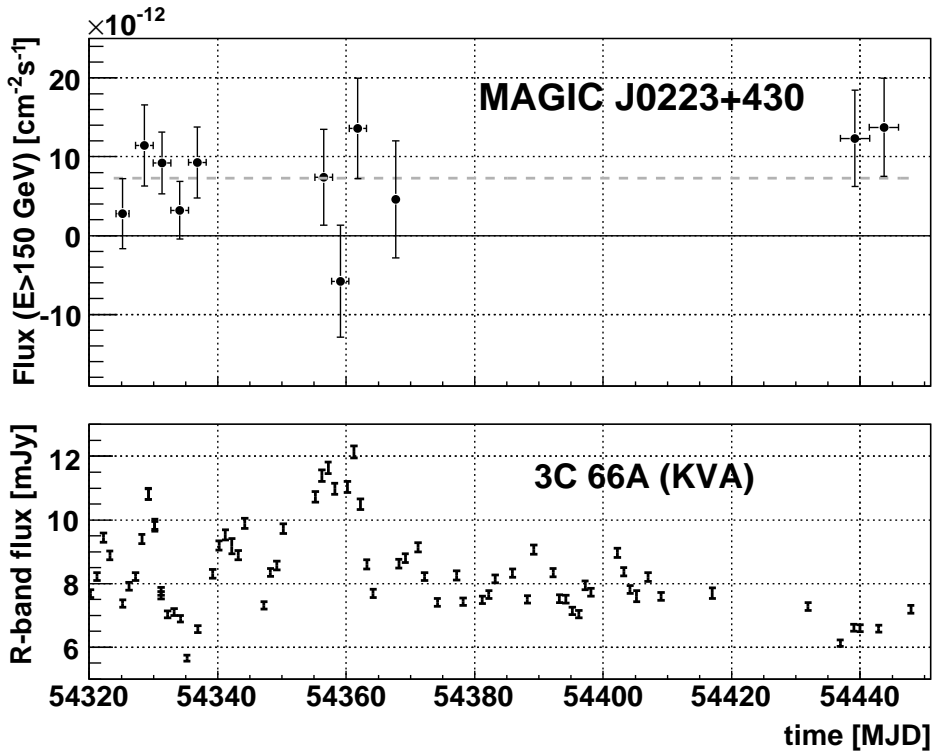} \label{sub_fig1}}
             }
   \caption{\textit{Upper panel:} MAGIC integral flux above 150\,GeV in bins of 3 days (except for periods where the sampling was coarser). The grey dashed line indicates the average $\gamma$-ray flux. \textit{Lower panel:} Optical light curve of 3C~66A as measured by the KVA telescope.
During the MAGIC observations 3C~66A was very bright at optical
wavelengths.
}
   \label{fig:lightcurve}
 \end{figure*}

Figure~\ref{fig:lightcurve} shows the light curve of MAGIC~J0223+430 together
with the flux of 3C~66A in optical wavelengths.  As we integrate over
$\gamma$-ray events from a wide sky region ($\sim0.07\, \mathrm{deg}^2$), we
cannot exclude that 3C~66A contributes to the measured signal.  The integral
flux above 150\,GeV corresponds to $\left(7.3\pm 1.5\right) \times
10^{-12}$\,cm$^{-2}$\,s$^{-1}$ (2.2\% of the Crab Nebula flux) and is the
lowest ever detected by MAGIC.  The $\gamma$-ray light curve is consistent with
a constant flux within statistical errors. These errors, however, are large,
and some variability of the signal cannot be excluded.

For the energy spectrum of MAGIC~J0223+430, loose cuts are made to keep the $\gamma$-ray acceptance high. The differential energy 
spectrum was unfolded and is shown in Figure~\ref{fig:spectrum}. The spectrum can be well fitted by a power law
which gives a differential flux (TeV$^{-1}$ cm$^{-2}$ s$^{-1}$) of:
\begin{equation}
\frac{\mathrm {d}N}{\mathrm{d}E\, \mathrm {d}A\, \mathrm {d}t} = (1.7\pm 0.3)\times10^{-11}(E/300\,\mathrm {GeV})^{-3.1\pm0.3} 
\end{equation}
The quoted errors are statistical only. The systematic uncertainty is estimated
to be 35\% in the flux level and 0.2 in the power law photon index \cite{crab}.
As we cannot exclude that 3C~66A contributes to the measured signal, the
spectrum shown in Figure~\ref{fig:spectrum} represents a combined
$\gamma$-ray spectrum from the observed region.


\section{Discussion and conclusions}
\label{conclusions}
A VHE $\gamma$-ray source MAGIC~J0223+430 was detected in 2007 August--December.
Given the position of the excess measured by MAGIC above 150\,GeV, the source of the $\gamma$-rays is most likely 3C~66B. The VHE $\gamma$-ray flux was found to be on the level of 2.2\% Crab Nebula flux and was constant during the observations.
The differential spectrum of MAGIC~J0223+430 has a photon spectral index of $\Gamma=3.10\pm 0.31$ and extends up to $\sim 2$\,TeV. 

\begin{figure}[tb]
\begin{center}
\includegraphics*[width=0.99\columnwidth]{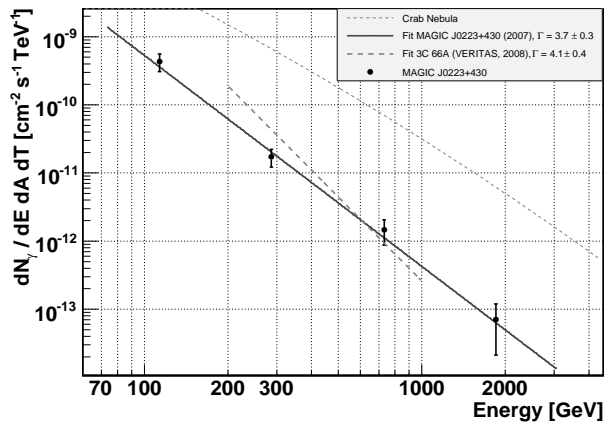}
\caption{
Differential energy spectrum of MAGIC~J0223+430. No correction for the $\gamma-\gamma$ attenuation due to the EBL has been made. The Crab Nebula spectrum \cite{crab} and the VERITAS spectrum on 3C~66A \cite{3c66averitas} is also shown as a reference (dashed grey line).
}
\label{fig:spectrum}
\end{center}
\end{figure}

The VERITAS collaboration recently detected VHE gamma-ray emission from 3C~66A \cite{3c66averitas}, with most of the signal coming from a gamma-ray flare 2008 September--October. The measured integral flux above 200\,GeV is 6\% of the Crab Nebula flux, and the energy spectrum is characterized by a soft power law with photon index $\Gamma = 4.1$ (see Figure~\ref{fig:spectrum}). In view of this detection, we note that if 3C~66A was emitting $\gamma$-rays in 2007 August to December then its flux was at a significantly lower level than in 2008.




The position of the source found by MAGIC is not spatially compatible with 3C~66A at a level of $1.5\sigma$. In addition to this discrepancy in the source position, the measured spectrum extending up to $\sim 2$\,TeV poses an additional drawback to the association of MAGIC~J0223+430 with 3C~66A.
Due to the energy-dependent absorption of VHE $\gamma$-rays with low-energy photons of the extragalactic background (EBL, \cite{gould}), the VHE $\gamma$-ray flux of distant sources is significantly suppressed.
We investigated the measured spectrum of MAGIC J0223+430 assuming a blazar-like emission mechanism, and following the prescription
of \cite{mazin}, deriving a redshift upper limit of the source to be $z<0.17$ ($z<0.24$) under the 
assumption that the intrinsic energy spectrum cannot be
harder than $\Gamma = 1.5$ ($\Gamma = 0.666$).
This assumption of $\Gamma > 1.5$ is based on particle acceleration
arguments \cite{AharonianEBL}, and the fact that none of the sources
in the EGRET energy band (not affected by the EBL) have shown a harder
spectrum, a trend confirmed by Fermi-LAT in their first three months of observations. The latter assumption of $\Gamma > 0.666$ can be considered as an extreme case of the spectrum hardness, 
suggesting a monochromatic spectrum of electrons when interacting with a soft photon target field \cite{katarzynski}.\footnote[2]{ 
See also \cite{stecker} for more detailed calculations.}
The derived redshift upper limit is incompatible with the association of MAGIC~J0223+430 with 3C~66A, unless the redshift of 3C~66A is much lower than the assumed value of $z=0.444$, or exotic scenarios are responsible for its VHE emission \cite{aha_abs}.

\newcommand{\degree}{{}^{\circ}}

3C~66B is a FRI radio galaxy similar to M~87, which has been detected to
emit VHE $\gamma$-rays \cite{hegraM87,m87}. 
Since the distance of 3C~66B is 85.5\,Mpc, its intrinsic VHE luminosity would be two to eight times
higher than the one of M~87 (22.5\,Mpc) given the reported variability of M~87 \cite{m87,MAGICm87}.

As in the case of M~87, there would be several possibilities for the region responsible of 
the TeV radiation in 3C~66B: the vicinity
of the supermassive black hole \cite{Neronov}, the unresolved base of the
jet (in analogy with blazar emission models; \cite{Tavecchio}) 
and the resolved jet. Unlike for M~87, we do not observe
significant variability in the VHE $\gamma$-ray flux and therefore we have
no constrains on the size of the emission region. However, as the
angle to line of sight is even larger than in M~87 (M~87: 19$\degree$,
\cite{Perlman}; 3C~66B: 45$\degree$, \cite{giovanni}) the resolved jet
seems an unlikely site of the emission. On the other hand, the unresolved base of the 
jet seems a likely candidate for the emission site as 
it could point with a smaller angle 
to the line of sight. 
If the viewing angle was small, blazar-like emission mechanisms cannot be
excluded. The orbital motion of 3C~66B shows evidence for a
supermassive black hole binary (SMBHB) with a period of
$1.05\pm0.03$ years \cite{sudou}. The SMBHB would likely cause the   
jet to be helical, and the pointing direction of the unresolved jet could
differ significantly from the direction of the resolved jet.

It has also been proposed that the spectrum obtained by MAGIC could be interpreted as combined emission from 3C~66B, dominating above $\sim 200$\,GeV, and 3C~66A \cite{tavecchio09}. The high-energy emission from the blazar would be strongly attenuated by the interaction with the EBL, and could only contribute to the measured excess at low energies. The emission from 3C~66B could be explained as radiation coming from the layer of the jet, in the framework of the structured jet model \cite{Ghisellini05} already used to interpret the TeV emission of M~87 \cite{Tavecchio}.

Given the likely association of MAGIC~J0223+430 with 3C~66B, our detection would 
establish radio galaxies as a new class of VHE $\gamma$-ray emitting sources.
According to \cite{Ghisellini05}, there are eight FR I radio galaxies in the 3CR 
catalog that should have a higher $\gamma$-ray flux at 100\,MeV than 3C 66B,
but possibly many of these 
sources are rather weak in the VHE $\gamma$-ray band. Further observations 
of radio galaxies with the Fermi Gamma-ray Space Telescope as well as by ground-based telescopes are needed 
to further study the $\gamma$-ray emission properties of radio galaxies.

\section*{Acknowledgments}
We thank the Instituto de Astrofisica de Canarias for the excellent working conditions at the 
Observatorio del Roque de los Muchachos in La Palma. The support of the German 
BMBF and MPG, the Italian INFN and Spanish MCINN is gratefully
acknowledged. This work was also supported by ETH Research Grant 
TH 34/043, by the Polish MNiSzW Grant N N203 390834, and by the YIP of the 
Helmholtz Gemeinschaft.


\end{document}